\documentclass{ws-procs9x6}

\begin{document}

\title{Challenges for Inflationary Cosmology 
}

\author{ROBERT H. BRANDENBERGER\footnote{\uppercase{W}ork partially
supported by (at \uppercase{M}c\uppercase{G}ill) an 
\uppercase{NSERC} \uppercase{D}iscovery \uppercase{G}rant and 
(at \uppercase{B}rown) by the 
\uppercase{US} \uppercase{D}epartment of \uppercase{E}nergy under 
\uppercase{C}ontract \uppercase{DE}-\uppercase{FG}02-91\uppercase{ER}40688, 
\uppercase{TASK A}.}}

\address{Physics Department, McGill University \\
3600 University Street \\ 
Montreal, PQ, H3A 2T8, CANADA\\
and\\ 
Physics Department, Brown University\\
Providence, RI 02912, USA\\
E-mail: rhb@hep.physics.mcgill.ca}

\maketitle

\abstracts{
Inflationary cosmology has provided a predictive and phenomenologically
very successful scenario for early universe cosmology. Attempts to implement
inflation using scalar fields, however, lead to models with serious conceptual
problems. I will discuss some of the problems, explain why string theory
could provide solutions to a subset of these problems, and give a brief 
overview of ``string gas cosmology'', one of the approaches to merge
string theory and early universe cosmology.}

\section{Introduction}

As is well known, inflationary cosmology was developed in order to
resolve some conceptual problems of standard big bang cosmology such
as the horizon and flatness problems \cite{Guth}. The most important success of
inflation, however, is that it provides a causal mechanism for
generating cosmological fluctuations on microscopic scales which
evolve into the perturbations which are observed in the large-scale
structure of the universe and the anisotropies in cosmic microwave background
(CMB) temperature maps \cite{flucts} (see also \cite{flucts2}). 
In the simplest models of inflation, models in which
inflation is driven by a single scalar field, the predicted spectrum
of fluctuations is adiabatic and almost scale-invariant. These predictions
have been confirmed with unprecedented accuracy by recent observations,
in particular by the WMAP maps of the temperature of the CMB \cite{WMAP}.

Before claiming that early universe cosmology is solved, we should
recall the status of the previous paradigm of early universe cosmology,
the ``standard big bang'' model. It explained the relation between
redshift and distance of galaxies. Without doubt, however, its biggest
success was that it predicted the existence and black body nature of
the cosmic microwave background. This spectacular quantitative success,
however, did not imply that the theory was complete. In fact, one of
the gravest conceptual problems of ``standard big bang'' cosmology
is related to its greatest successes: the theory does not explain why
the temperature of the black body radiation is nearly isotropic across
the sky. 

In the following I will argue that the calculations involved in the
derivation of the spectrum of cosmological fluctuations contain in
themselves the seeds for their incompleteness. Other conceptual problems
of standard big bang cosmology were not resolved by inflation and thus
re-appear in the list of conceptual problems of inflationary cosmology. 

Most models of inflation are formulated in the context of Einstein
gravity coupled to a matter sector which contains the ``inflaton'',
a new scalar field. Some (but not all) of the problems discussed below
are specific to such scalar field-driven models of inflation. I will
focus on this class of inflationary models.

\section{Conceptual Problems of Scalar Field Driven Inflation}

\subsection{Fluctuation Problem}

As was mentioned in the Introduction, inflationary cosmology produces
an almost scale-invariant spectrum of cosmological fluctuations \cite{flucts}.
A concrete model of scalar field-driven inflation also predicts the
amplitude of the spectrum. The problem is that without introducing
a hierarchy of scales into the particle physics model, the predicted
amplitude of fluctuations exceeds the observational results by several
orders of magnitude. For example, if the potential of the inflaton (the
scalar field driving inflation) $\varphi$ has the form
\begin{equation}
V(\varphi) \, = \, {1 \over 4} \lambda \varphi^4 \, ,
\end{equation}
then a value of 
\begin{equation} \label{hierarchy}
\lambda \, \sim \, 10^{-12}
\end{equation}
is required in order that the predicted amplitude of the spectrum matches
with observations. It has been shown \cite{Adams} that this problem
is quite generic to scalar field-driven inflationary models (counterexamples,
however, can be constructed - I thank David Lyth for pointing this out to
me). Given that
one of the main goals of cosmological inflation is to avoid fine tunings
of parameters of the cosmology, it is not very satisfying that fine tunings
re-appear in the particle physics sector.

\subsection{Trans-Planckian Problem}

The most important success of inflationary cosmology is that scales
of cosmological interest today start out with a wavelength smaller
than the Hubble radius at the beginning of the period of inflation. This
allows the development of a causal theory of the generation and evolution
of fluctuations (see e.g. \cite{MFB} for a comprehensive review, and
\cite{RHBrev2} for a shorter summary). In most scalar field-driven inflationary
models, in particular those of chaotic \cite{chaotic} 
type, the period of inflation lasts much longer
than the minimal number of e-foldings required for successful resolution
of the horizon and flatness problems. In this case, the physical wavelength
of comoving modes which correspond to today's large-scale structure
was smaller than the Planck length at the time $t_i$ when the period of
inflation began (see Fig. 1 (taken from \cite{Jerome1}) 
for a sketch of the space-time geometry).

It is clear that both key ingredients of scalar field-driven inflationary
cosmology, namely the description of space-time by General Relativity, and
the use of semi-classical scalar fields to model matter, break down on scales
smaller than the Planck length. However, the spectrum of cosmological
fluctuations is determined by solving the equations for perturbations as
an initial value problem, starting off all modes at time $t_i$ in their
vacuum state. The unknown physics which describes space-time and matter
on Planck scales must enter into the calculation of the spectrum of
cosmological fluctuations. The key question \cite{RHBrev1} is whether
the conclusions derived from the usual computations are sensitive to the
new physics.

\begin{figure}[ht]
\centerline{\epsfxsize=4.1in\epsfbox{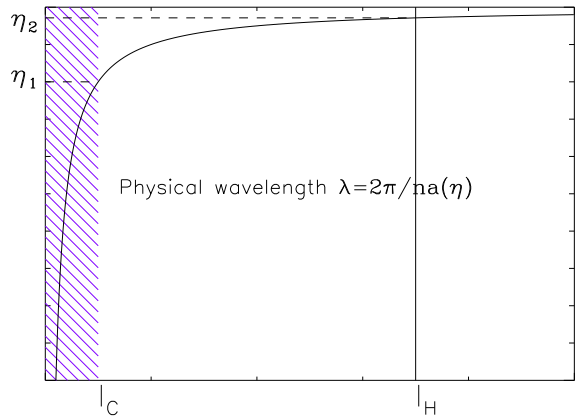}}   
\caption{Space-time sketch showing how the physical wavelength (horizontal
axis) of a fluctuation with fixed comoving frequency $n$ increases in
time (vertical axis) during the period of inflation. The variable $\eta$
denotes conformal time, the length $l_c$ is the Planck length (or whatever
the critical ultraviolet length of the new physics is), $l_H$ denotes
the Hubble radius.}
\end{figure}

Let us give an argument why large deviations in the predictions could
be expected \cite{Jerome1}. Returning to Fig. 1, we see that the
evolution of the perturbations for any mode can be divided into three
phases. In the first phase ($\eta < \eta_1$), 
the wavelength is smaller than the Planck
length and the evolution is crucially influenced by the new physics.
In Period 2 ($\eta_1 < \eta < \eta_2$), 
the wavelength is larger than the Planck length but smaller
than the Hubble length. In this period, the fluctuations will perform
quantum vacuum oscillations, as described by the usual theory of
cosmological fluctuations \cite{MFB,RHBrev2}. Finally, in the third
period (wavelength larger than the Hubble scale), the fluctuations
freeze out, are squeezed, and become classical. Assume now that the
evolution in Phase 1 is such that it is non-adiabatic from the point
of view of the usual equations. In this case, an initial state at $t_i$
minimizing the energy (in the frame given by the cosmological background)
will evolve into an excited state at the time $t_c(k)$ when the mode
$k$ enters Period 2. Since different modes spend a different amount of
time in Period 1, one should expect that the excitation level at time
$t_c(k)$ will depend on $k$ (in fact, a blue spectrum will be produced),
in contrast to what happens in the usual analysis (which yields a
scale-invariant spectrum). A toy model of new physics which gives
this result was presented \cite{Jerome1}. It is based on describing
the evolution of fluctuations by means of a modified dispersion
relation $\omega_p(k_p)$, where the subscripts $p$ indicate physical
(rather than comoving) values.

Obviously, more conservative choices for describing the evolution
of fluctuations in the presence of new physics, e.g. the ``minimal
trans-Planckian'' prescription of starting the perturbation modes
off in a vacuum state at the time that the wavelength equals the
scale of new physics \cite{Ulf,Ven1} yield corrections to the
usual predictions which are small (see e.g. \cite{Jerome2} for
a review with references to original works). It was suggested 
\cite{Tanaka,Starob} that back-reaction effects of ultraviolet modes might
lead to stringent limits on the amplitude of possible
trans-Planckian effects. However, it was recently shown \cite{Jerome3}
that the back-reaction of ultraviolet modes in fact mainly leads
to a renormalization of the cosmological constant, and that the
constraints on the magnitude of possible trans-Planckian effects which
are consistent with having a period of inflation are thus greatly
relaxed.

The ``trans-Planckian'' problem for inflationary cosmology should
in fact not be viewed as a problem at all. Rather, it opens an exciting
window to probe Planck-scale physics using current cosmological 
observations.

\subsection{Singularity Problem}

The famous Hawking-Penrose theorems \cite{HE} prove that
standard big bang cosmology is incomplete since there is an initial
singularity. These theorems assume that matter has an equation of
state with non-negative pressure, and that space-time is described by
the Einstein equations. As was shown recently \cite{Borde}, these theorems
can be generalized to apply to scalar field-driven inflationary
cosmology. It is shown that there are some geodesics which cannot
be extended arbitrarily far into the past. An intuitive way to
understand this result is the following. We focus on some time $t$
during the phase of inflation. Some fluctuations will be present at
this time. Evolving these fluctuations back in time, we find that
they will come to dominate the energy density at some time in the
past. Prior to this time, the equation of state of scalar field
matter thus has non-negative pressure, and the usual theorems apply.
It thus follows that scalar field-driven inflationary cosmology
is incomplete in the same way that standard big bang cosmology
was incomplete.

\subsection{Cosmological Constant Problem}

The Achilles heel of scalar field-driven inflationary cosmology is
the fact that the scenario uses the time-independent part of the
potential energy of the scalar field to generate inflation.
However, it is known that another form of constant potential energy,
namely the quantum vacuum energy, does not gravitate. There is
some unknown mechanism (the solution to the famous cosmological
constant problem) which will explain why the quantum vacuum energy
does not gravitate.
The key challenge is to show why this mechanism does not also
render the constant part of the scalar field potential energy
gravitationally inert.

If the cosmological constant problem is solved by gravitational
back-reaction (see \cite{RHBrev3} for a review of this scenario),
then scalar field-driven inflation would be robust. However, this
scenario \cite{RHBrev3} is at the moment rather controversial
and requires much more development.

\subsection{Who is the Inflaton?}

The initial hope \cite{Guth} was that the inflaton could be identified
with the Higgs field of particle physics. However, because of hierarchy
problems like (\ref{hierarchy}), this hope was not realized.
At the present time, although there are many possible models
of inflation, there are no convincing theories based on well-established
particle physics. Thus, an outstanding challenge for inflationary
cosmology is to determine what the origin of the inflaton field is.

\subsection{Why String Theory May Help}

All of the above conceptual problems of current inflationary cosmology
stem from our incomplete understanding of the fundamental physics at
ultra-high energies. String theory is a candidate for a unified quantum
theory of space-time and matter. Thus, it is challenging to study
whether string theory provides a framework to possibly resolve the
abovementioned problems.

Since string theory contains many scalar fields which are massless before
supersymmetry breaking, it provides both candidates for the inflaton and
an obvious way of producing the
hierarchy of scales required to resolve the fluctuation problem. In the case
of low-scale supersymmetry breaking, both the slow-rolling conditions and
the size of the mass hierarchy might be sufficient for a successful inflationary
model (I again thank David Lyth for stressing to be the caveat of
low-scale supersymmetry breaking). Since
string theory should describe the physics on all length scales, string
theory should provide the correct equation to describe cosmological
fluctuations throughout the evolution of the universe, thus resolving
the trans-Planckian problem. Finally, it is hoped that string theory
will resolve cosmological singularities (and a concrete framework in
which this happens exists \cite{Vafa}). The only of the conceptual problems
of inflation which at the present time does not appear
to be solvable within the current knowledge of 
string theory is the cosmological constant problem.

\section{Challenges for String Cosmology}

However, when considering string theory as a framework for early universe
cosmology, an immediate obstacle appears: at a perturbative level,
critical superstring theory is consistent only in nine spatial dimensions.
Why do we only see three spatial dimensions? There is evidence that
the non-perturbative formulation of string theory (``M-theory'') in
a different limit leads to 11-d supergravity. Once again, the predicted
number of spatial dimensions is not what is observed.

The traditional approach to resolving this problem is to assume that
six (or seven) of the spatial dimensions are compactified on a string-scale
manifold and hence are invisible to us. From the point of view of
cosmology there are two immediate questions: First, what selects the
number of dimensions which are compactified, and second, why is the radius
of compactification stable? In a more modern approach (``brane world
scenarios''), the matter fields of the particle physics standard model
are taken to be confined to a 3-brane (three spatial dimensions) which lives
in the higher-dimensional bulk. Again, however, the question as to why
a 3-brane (and not a brane of different dimensionality) is the locus 
of our matter fields arises (there has been an interesting work on this issue
\cite{Mahbub}).
 
Thus, before addressing the issue of whether string theory can provide
a convincing realization of cosmological inflation, string cosmology
should explain why there are only three large spatial dimensions. The
next section will briefly review a string-based scenario of the early
universe which may provide an explanation. Once three spatial dimensions
have been selected as the only ones to become large, string theory
offers various mechanisms of obtaining inflation
\cite{Dvali,KKLMMT}. 

\section{Overview of String Gas Cosmology}

A frequent criticism of current approaches to string cosmology is that
string theory is not developed enough to consider applications to
cosmology. In particular, one may argue that string cosmology must be
based on a complete non-perturbative formulation of string theory. 
By focusing on the effects of a new symmetry of string theory, namely
t-duality, and of truly stringy degrees of freedom, namely string
winding modes, the ``string gas cosmology'' program initiated in 
\cite{Vafa} (see also \cite{Perlt}) 
and resurrected \cite{ABE} after the D-brane ``revolution'' in string 
theory  hopes to make predictions which will also be
features of cosmology arising from a non-perturbative string theory.  

In string gas cosmology it is assumed that, at some initial time, the
universe begins small and hot. For simplicity, we take space to be a
torus $T^9$, with all radii $R$ equal and of string length. The t-duality
symmetry - which is central to the scenario - is the invariance of the spectrum
of free string states under the transformation 
\begin{equation} \label{dual1}
R \, \rightarrow \, 1/R
\end{equation}
(in string units). String states consist of center of mass momentum modes
whose energy values are quantized in units of $1/R$, winding modes whose
energy values are quantized in units of $R$, and oscillatory modes whose
energy is independent of $R$. If $n$ and $m$ denote the winding and momentum
numbers, respectively, then the spectrum of string states is invariant under
(ref{dual1}) if
\begin{equation}
(n, m) \, \rightarrow \, (m, n) \, .
\end{equation}
T-duality is also a symmetry of non-perturbative string theory \cite{Pol}.

The action of the theory is given by the action of a dilaton-gravity
background (a background for which the action has t-duality symmetry in
contrast to what would be the case were we to fix the dilaton and consider
the Einstein action only) coupled to a matter action describing an ideal
gas of all string matter modes. In analogy to standard big bang cosmology
we assume that at the initial time all matter modes (in particular string
momentum and winding modes) are excited. We call these initial conditions
``democratic'' - since all radii are taken to be equal - and ``conservative''
- since hot rather than cold initial conditions are assumed. If we start
from the Type II superstring corner of the M-theory moduli space, there
will be branes of various dimensionalities in addition to the fundamental
strings.

Using the dilaton gravity equations of motion is can be shown \cite{Vafa2,Ven2}
that negative pressure leads to a confining potential for the radius of
the tori. Thus, the presence of string winding modes will prevent the spatial
dimensions from expanding. String momentum modes will not allow space to
contract to zero size. There will be a preferred radius (the ``self-dual''
radius which in string units is $R = 1$). Spatial dimensions can only become
large if the string winding modes can annihilate (we assume zero net winding
number). This cannot occur in more than three large dimensions because the
probability for string world sheets to intersect will be too small. In three
dimensions, as studied in detail elsewhere \cite{BEK}, the winding modes will
fall out of equilibrium as $R$ increases, and they
can annihilate sufficiently fast to allow these three dimensions to expand
without bounds. Thus, in the Type II corners of the M-theory moduli space,
string gas cosmology provides a possible solution of the dimensionality problem
facing any string cosmology. By t-duality, a universe with $R < 1$ is
equivalent to a universe with $R^{\prime} = 1/R > 1$. Hence \cite{Vafa}, 
string gas cosmology also provides a nonsingular cosmological scenario.

Note that the existence of fundamental strings or stable 1-branes is crucial
to having $D = 3$ emerge dynamically as the number of large spatial
dimensions. In the 11-d supergravity corner of the M-theory moduli space
it in unlikely \cite{Columbia1} that $D = 3$ will emerge as the number of
large spatial dimensions. There are, however, possible loopholes 
\cite{Stephon2} in this conclusion based on the special role which 
intersections of D=2 and D=5 branes can play.

Once three spatial dimensions start to expand, the radius of the other
dimensions (the ``radion'') will automatically be stabilized at the
self-dual radius \cite{Watson}. This conclusion is true for a background
given by dilaton gravity. To make contact with the late-time universe,
a mechanism to stabilize the dilaton must be invoked. It has been shown
\cite{Patil} that, at least in the case of one extra spatial dimension,
the radion continues to be stabilized after the dilaton is fixed. Crucial
to reach the conclusion \cite{Patil,WatBatt,Wat} is the inclusion in the 
spectrum of string states of states which become massless at the self-dual 
radius.

Thus, string gas cosmology appears to provide a scenario which provides
a nonsingular cosmology, explains why D=3 is the number of large spatial
dimensions at late times, and incorporates radion stabilization without
any extra physics input. The two main challenges to the scenario are
to provide a mechanism for dilaton stabilization and to provide a solution
to the flatness problem. It would be nice if string gas cosmology would
provide a natural mechanism for inflating the three large dimensions. There
are some initial ideas \cite{BGinflation} towards this goal.


\end{document}